\documentclass{ws-procs9x6}

\begin{document}

\title{The quark-quark correlator: theory and phenomenology}

\author{Elvio Di Salvo}

\address{Dipartimento di Fisica, Universita' di Genova - INFN, sez. Genova\\
Via Dodecaneso 33\\ 
16146 Genova - Italy\\ 
E-mail: disalvo@ge.infn.it}

\maketitle

\abstracts{
New properties of the quark correlator are found via equations of motion. As a 
first result, approximate relations can be established among the "soft" distribution 
functions; one such relation may help in determining the quark transversity in a nucleon. 
Secondly, the $Q^2$ dependence of the T-odd functions is deduced; the result is compared to
unpolarized Drell-Yan data. Lastly, important remarks are made about the contributions to 
$g_2(x)$.}

\section{Introduction}
$~~~~$ The correlator, a very important theoretical tool encoding the "soft" functions 
involved in high energy reactions, was originally introduced by Ralston and Soper in 
1979\cite{rs} and successively adopted by other authors\cite{mt,bjm,gms}, who studied some 
properties of this matrix. However, as I shall show, further results can be established in 
the sector, thanks to the equations of motion. These new results have important 
phenomenological consequences. In particular, I point out that progress can be done about 
three problems which arise in high energy physics, {\it i. e.}, determining transversity, 
interpreting azimuthal asymmetries and disentangling the contributions to $g_2(x)$.

a) It is not easy to determine experimentally the transversity in a spinning hadron, because 
this density is chiral odd and has to be coupled with another chiral odd function, for 
example the transversity of an antiquark in doubly polarized Drell-Yan, or the Collins 
function (or the Jaffe interference function) in singly polarized semi-inclusive deep 
inelastic scattering (SIDIS). The various methods proposed for determining this function 
present more or less serious drawbacks.

b) Azimuthal asymmetries of surprisingly large size have been observed in unpolarized 
Drell-Yan, in singly polarized SIDIS and in inclusive hadronic reactions. Such asymmetries
do not find an explanation in ordinary perturbative QCD. Among possible interpretations, we 
recall quark-quark-gluon correlations\cite{qs} and, more recently, the T-odd 
functions\cite{bjm,bhs}, which take into account the intrinsic transverse momentum of a quark 
inside a hadron. These functions provide a discrete description of the azimuthal asymmetries, 
however, as regards the $Q^2$ dependence of such asymmetries, they do not agree neither with 
the predictions of quark-quark-gluon correlations, nor with unpolarized Drell-Yan 
data\cite{fa}.

c) The function $g_2(x)$ has been studied by several authors\cite{ael,jj} from various 
viewpoints, but there is no agreement about the contributions it involves. 

In my talk I shall show that the equations of motion\cite{po} allow to establish for the  
correlator some new properties, which lead to partial answers to the three problems just 
illustrated. First of all, I shall define the correlator. Secondly, I shall split it into a 
T-even and a T-odd part. Thirdly, I shall expand it in powers of the coupling and I shall 
apply the equations of motion, getting some conditions on the first three terms of the 
expansion. I shall discuss each such term in some detail. Lastly, I shall draw a short 
conclusion.

\section{The Correlator}

The correlator is defined as\cite{mt}  
\begin{equation}
\Phi_{ij}(p; P_0,S) = \int\frac{d^4x}{(2\pi)^4} e^{ipx} 
\langle P_0,S|\bar{\psi}_j(0)  L(x)  \psi_i(x)|P_0,S\rangle. \label{corr}
\end{equation}
Here $\psi$ is the quark field, $p$ the quark four-momentum and $|P_0,S\rangle$ denotes a 
nucleon state with a given four-momentum $P_0$ and Pauli-Lubanski four-vector $S$. Moreover
$L(x)$ is the gauge link operator\cite{mt}, {\it i. e.},
\begin{equation}
 L(x) = {\mathrm P} exp\left[-ig\Lambda_P(x)\right], \ ~~~~~~ \ 
\Lambda_P(x) = \int_0^x \lambda_a A^a_{\mu}(z)dz^{\mu}. \label{phase}
\end{equation}
Here "P" denotes the path-ordered product along the integration contour $P$; $g$, $\lambda_a$ 
and $A^a_{\mu}$ are respectively the strong coupling constant, the Gell-Mann matrices and the 
gluon fields. The link operator depends on the choice of $P$, which has to be fixed so as to 
make a physical sense\cite{co}. According to previous treatments\cite{co,mt}, I define two 
different contours, $P_{\pm}$, as sets of three pieces of straight lines, from the origin to 
$x_{1\infty}\equiv (\pm\infty, 0, {\bf 0}_{\perp})$, from $x_{1\infty}$ to $x_{2\infty}\equiv 
(\pm\infty, x^+, {\bf x}_{\perp})$ and from $x_{2\infty}$ to  $x\equiv (x^-, x^+,{\bf 
x}_{\perp})$; here the + or - sign has to be chosen, according as to whether final or initial 
state interactions\cite{co} are involved in the reaction. I have adopted a frame - to be 
used throughout this talk - whose $z$-axis is taken along the nucleon momentum, with $x^{\pm} 
= 1/\sqrt{2}(t\pm z)$. 

The correlator enjoys two important properties, due to the hermiticity condition and 
parity conservation:   
\begin{equation}
\Phi^{\dagger} = \gamma_0 \Phi \gamma_0,  \ ~~~~~~ \ ~~~   \Phi(p,P_0,S) = \gamma_0 
\Phi({\bar p},{\bar P}_0,-{\bar S}) \gamma_0.
\end{equation}
Here ${\bar p} \equiv (p_0, -{\bf p})$, having set $p \equiv (p_0, {\bf p})$; ${\bar P}_0$ 
and ${\bar S}$ are 
defined analogously. Time reversal invariance does not give rise to any condition on $\Phi$.
Indeed, we may have T-even and T-odd functions, the latter ones being generated by 
interference between two amplitudes which behave differently under time reversal. 

\section{Splitting}

I set
\begin{equation}
\Phi_{E(O)} = \frac{1}{2}[\Phi_+\pm\Phi_-], \label{spl}
\end{equation}
where $\Phi_{\pm}$ corresponds to the contour $P_{\pm}$ in eqs. (\ref{phase}), while
$\Phi_E$ and $\Phi_O$ select respectively the T-even and the T-odd "soft" functions. These 
two correlators contain respectively the link operators $L_E(x)$ and $L_O(x)$, where
\begin{equation}
L_{E(O)}(x) = \frac{1}{2} {\mathrm P} \left\{exp\left[-ig\Lambda_{P_+}(x)\right]\pm 
exp\left[-ig\Lambda_{P_-}(x)\right]\right\} \label{spl1}
\end{equation}
and $\Lambda_{P_{\pm}}(x)$ are defined by the second eq. (\ref{phase}). Notice that, 
for T-even functions, the result is independent of the contour ($P_+$ or $P_-$), while T-odd 
functions change sign according as to whether they are generated by 
initial or final state interactions\cite{co}. In this sense, such functions are not strictly 
universal\cite{co}. Eq. (\ref{spl1}) has important consequences on $\Phi$ for small values of 
$g$. Indeed, the zero order term is just T-even, while the first order correction contains 
T-odd contributions. This confirms that no T-odd terms occur without interactions among 
partons, as claimed by other authors\cite{bhs,co}. 

An immediate advantage of the splitting (\ref{spl}) is that we can define separately T-even 
and T-odd functions, by projecting $\Phi_E(x)$ and $\Phi_O(x)$  over the various Dirac 
components. These projections are defined as
$\Phi^{\Gamma} = \frac{1}{2} \int dp^- tr(\Phi\Gamma)$,
where $\Gamma$ is a Dirac operator. In particular, among the T-even functions, I consider the 
three main densities of quarks in the nucleon, {\it i.e.}, the unpolarized density 
$f_1(x,{\bf p}_{\perp}^2)$, the longitudinally polarized density $g_{1L}(x,{\bf 
p}^2_{\perp})$ and the transversity $h_{1T}(x,{\bf p}_{\perp}^2)$. They are related, 
respectively, to $\Phi_E^{\gamma^+}$, to $\Phi_E^{\gamma_5\gamma^+}$ and to 
$\Phi_E^{\gamma_5\gamma^+\gamma^i}$, $i = 1,2$. Among the T-odd functions, it is worth 
mentioning the unpolarized quark density $f_{1T}^{\perp}(x,{\bf p}_{\perp}^2)$ in a 
transversely polarized nucleon (the Sivers\cite{si} function) and the transversity 
$h_{1}^{\perp}(x,{\bf p}_{\perp}^2)$ in an unpolarized nucleon (analogous to the 
Collins\cite{co1} fragmentation function), which may be derived, respectively, from 
$\Phi_O^{\gamma^+}$ and $\Phi_O^{\gamma_5\gamma^+\gamma^i}$, $i = 1,2$.

Now I consider a particular gauge, that is, {\it an axial gauge with antisymmetric boundary 
conditions}\cite{mt}. In this case one has $\Lambda_{P_-}(x) = -\Lambda_{P_+}(x)$ and 
therefore
\begin{equation}
L_E(x) = {\mathrm P} cos\left[g\Lambda_{P_+}(x)\right], \ ~~~~~~ \
L_O(x) = -i{\mathrm P} sin\left[g\Lambda_{P_+}(x)\right].
\end{equation}
In this particular gauge - to be referred to as $G$-gauge  in the following - the T-even 
(T-odd) part of the correlator consists of a series of even (odd) powers of $g$, each term 
having an even (odd) number of gluon legs.

\section{Equations of motion}
 
Now I invoke the Politzer theorem on equations of motion\cite{po}, {\it i.e.}, 
\begin{equation}
\langle P_0,S|F(\psi) (iD\hspace{-0.7em}/-m_q)\psi(x)|P_0,S\rangle = 0. \label{pol}
\end{equation}
Here $m_q$ is the quark rest mass, $F(\psi)$ a functional of the quark field and 
$D_{\mu} = \partial_{\mu} - ig \lambda_a A^a_{\mu}$ the covariant derivative. The result 
(\ref{pol}) survives renormalization. I adopt the $G$-gauge  and expand the correlator in  
powers of $g$, {\it i. e.},
\begin{eqnarray}
\Phi_E(p) &=& \Phi^{(0)}_E(p) + g^2\Phi^{(2)}_E(p) + O(g^4), \label{even}
\\
\Phi_O(p) &=& g\Phi^{(1)}_O(p) + O(g^3).
\end{eqnarray}
Setting $F(\psi)$ = ${\bar \psi(0)}L(x)$, eq. (\ref{pol}) yields, after some steps\cite{ds}, 
\begin{eqnarray}
(p\hspace{-0.45em}/-m_q)\Phi^{(0)}_E(p) &=& 0,
\\
(p\hspace{-0.45em}/-m_q)\Phi^{(1)}_O(p) &=& \Psi_O(p),
\\
(p\hspace{-0.45em}/-m_q)\Phi^{(2)}_E(p) &=& \Psi_E(p).
\end{eqnarray}
Here 
\begin{eqnarray}
\Psi_O &=& \Psi[{A\hspace{-0.60em}/}_2-{A\hspace{-0.60em}/}]\label{odd0}
\\
\Psi_E &=&\Psi[A_2^2]-(p\hspace{-0.45em}/-m_q)
\Psi[{A\hspace{-0.60em}/}_1{A\hspace{-0.60em}/}](p\hspace{-0.45em}/-m_q)^{-1}\label{noti},
\end{eqnarray}

${A\hspace{-0.60em}/} = \lambda_a {A\hspace{-0.60em}/}^a(x)$, ${A\hspace{-0.60em}/}_i = 
\lambda_a {A\hspace{-0.60em}/}^a(x_{i\infty})$ (i = 1, 2), $A_2^2 = {A\hspace{-0.60em}/}_2 
{A\hspace{-0.60em}/}_2$ and
\begin{equation}
\{\Psi[N]\}_{ij} = \int\frac{d^4x}{(2\pi)^4} e^{ipx} \langle P_0,S|\bar{\psi}_j(0)N_{ik} 
\psi_k(x)|P_0,S\rangle
\end{equation}
is a functional of the generic Dirac operator $N$. 
As a consequence of eqs. (\ref{even}) to (\ref{noti}), the expansions of $\Phi_E(p)$ and 
$\Phi_O(p)$ in powers of $g$ read\cite{ds}
\begin{eqnarray}
\Phi_E(p) &=& p^+ \rho \delta(p^2-m_q^2) + g^2\kappa^2_{\perp}M_E(p) + O(g^4), \label{even1}
\\
\Phi_O(p) &=& g\kappa_{\perp} M_O(p) + O(g^3), \label{odd1}
\end{eqnarray}
where 
\begin{equation}
\rho = \frac{1}{2}(p\hspace{-0.45em}/+m_q)(f_1+\gamma_5{S\hspace{-0.65em}/}^q_{\parallel}
g_{1L}+\gamma_5{S\hspace{-0.65em}/}^q_{\perp}h_{1T}), \label{dmat}
\end{equation}
\begin{equation}
M_{E}\kappa^2_{\perp} = (p\hspace{-0.45em}/-m_q)^{-2}\Psi_E, \ ~~~~~~ \ 
M_{O}\kappa_{\perp} =  (p\hspace{-0.45em}/-m_q)^{-1}{\Psi_O} \label{psis}
\end{equation}
and
\begin{equation}
\kappa_{\perp} = \frac{|{\bf p}_{\perp}|}{p^+}. \label{kgre}
\end{equation}
$S^q_{\parallel}$ and $S^q_{\perp}$ are respectively the longitudinal and transverse 
components of the Pauli-Lubanski four-vector of the quark.
Notice that four-momentum conservation in the correlators (\ref{odd0}) and (\ref{noti}) 
avoids singularities in eqs. (\ref{psis}).

\section{Consequences and comments}

~~ A) At zero order in $g$ the correlator reduces essentially to the density matrix $\rho$ 
[eq. (\ref{dmat})] of a free, on-shell quark. Since $\rho$ depends just on three functions, 
one deduces several approximate relations among the functions involved in the parametrization 
of the correlator\cite{mt,bjm}. In particular, we have, in the $G$-gauge, 
\begin{eqnarray}
h_{1T}(x,{\bf p}^2_{\perp}) &=& g_{1T}(x,{\bf p}^2_{\perp}) + O(\alpha_s\kappa^2_{\perp}), 
\label{h1}\\
h_{1L}(x,{\bf p}^2_{\perp}) &=& g_{1L}(x,{\bf p}^2_{\perp}) + O(\alpha_s\kappa^2_{\perp}).
\label{g1} 
\end{eqnarray}
Here $g_{1T}$ is the helicity density of a quark in a transversely polarized nucleon, while 
$h_{1L}$ is the transversity of a quark in a longitudinally polarized nucleon. Relations 
(\ref{h1}) and (\ref{g1}) are especially important, because they relate chiral odd functions 
to chiral even ones, which may be determined much more easily\cite{ds1}. Such relations are 
not altered by QCD perturbative evolution, since the Politzer theorem survives 
renormalization. Therefore eq. (\ref{h1}) is useful in determining transversity. However, 
such a relation, as well as eq. (\ref{g1}), may be slightly modified by terms of order 
$O(\alpha_s\kappa^2_{\perp})$ and also by other contributions, as we shall see in a moment.

B) Eq. (\ref{odd1}) implies that first order corrections in the correlator are of order 
$O(g\kappa_{\perp})$; this conclusion turns out to be gauge independent\cite{ds}. Moreover 
the asymmetries in SIDIS and Drell-Yan which involve T-odd functions are suppressed at least 
as $Q^{-n}$, where $Q$ is the hard QCD scale and $n$ is the number of T-odd functions 
involved. This finds an experimental confirmation in the azimuthal asymmetry of unpolarized 
Drell-Yan\cite{fa}. Indeed, in the formalism of T-odd functions, this asymmetry is 
proportional to the convolutive product $h_1^{\perp}\otimes \bar{h}_1^{\perp}$\cite{bbh}, 
therefore, according to our predictions, it decreases as $Q^{-2}$. This result is confirmed 
by the best fits to data\cite{fa}, see figs. 1 and 2. Lastly, eq. (\ref{odd0}) shows that 
T-odd terms are produced by quark-quark-gluon correlations, which may be approximated by 
"effective" quark distributions only under some particular conditions\cite{co}. 

\begin{figure}[ht]
\centerline{\epsfxsize=4.1in\epsfbox{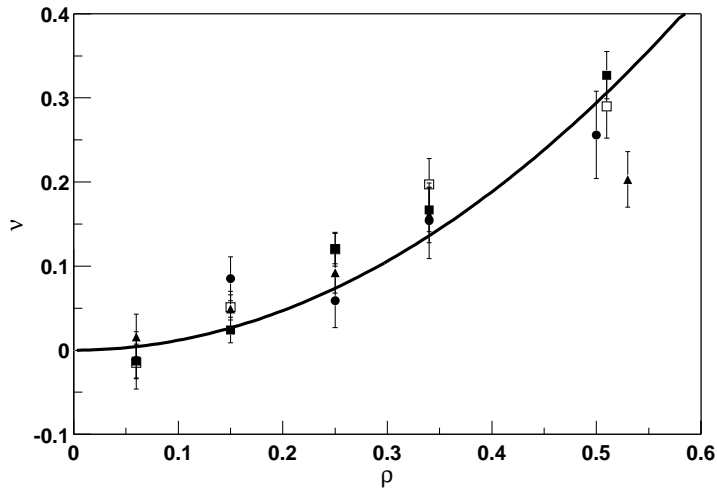}}   
\caption{Unpolarized Drell-Yan (see refs. 7): the asymmetry parameter $\nu$ {\it vs} $\rho = 
|{\bf q}_{\perp}|/Q$, where ${\bf q}_{\perp}$ and $Q$ are respectively the transverse 
momentum and the effective mass of the muon pair.}
\end{figure}

\begin{figure}[ht]
\centerline{\epsfxsize=4.1in\epsfbox{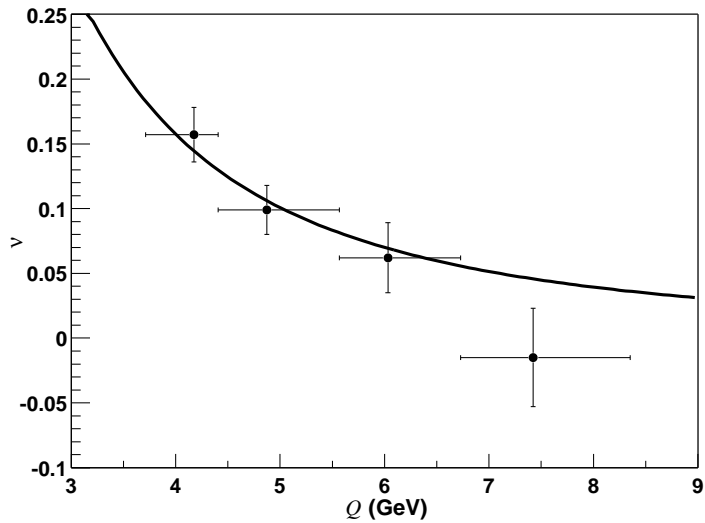}}   
\caption{Unpolarized Drell-Yan (see refs. 7): the asymmetry parameter $\nu$  {\it vs} $Q$ at 
fixed $|{\bf q}_{\perp}|$. Same notations as in fig. 1.}
\end{figure}  
C) Other important consequences can be drawn from  eq. (\ref{even1}) in the $G$-gauge and in 
the limit of small $g$. In particular, as regards the functions $g_1$ and $g_2$, I derive
\begin{eqnarray}
g_T = g_1 + g_2 = (m_q/xM) h_1 + O[\alpha_s(\pi_{\perp}/p^+)^2],\label{g2}
\end{eqnarray}
where $M$ and $\pi_{\perp}$ are, respectively, the nucleon rest mass and the mean value of 
$|{\bf p}_{\perp}|$. This relation is gauge invariant. It does not include the contribution 
of the anomalous coupling of the singlet axial current with gluons, nor of nonperturbative 
fluctuations of the nucleon, like, {\it e. g.}, $N\to \Delta \pi$\cite{sg}: indeed, the 
latter term demands a large value of $g$, making approximation (\ref{even}) meaningless. 
Therefore nontrivial twist-2 and twist-3 contributions to the combination $g_1+g_2$, which 
are allowed by the operator product expansion\cite{ael}, may arise only from these two terms. 
Incidentally, contributions from nonperturbative fluctuations are of order 0.1$g_1$\cite{sg} 
and may, in principle, affect by a few percent all the Dirac components 
of the correlator (\ref{even}), and therefore also eqs. (\ref{h1}) and (\ref{g1}). 

Moreover, one has $\sqrt{2}p^+ \simeq Q$ in processes like Drell-Yan and SIDIS; therefore 
eqs. (\ref{g2}) and (\ref{kgre}) imply that the main power correction to $g_T$ is of order 
$\alpha_s(\pi_{\perp}/Q)^2$. This correction comes from terms of the type (\ref{noti}), with 
2 gluon legs. Therefore the $\gamma_5\gamma^i$ projection ($i = 1,2$) of the 
quark-quark-gluon correlator [eq. (\ref{odd0})] trivially fulfills the Burkhardt-Cottingham 
sum rule\cite{bc}, in accord with the result of ref. 19. 

\section{Conclusions}

Here I shortly recall the main results exposed in my talk, which cast a new light on the 
problems illustrated in the introduction.
 
- I establish approximate relations between chiral-even and chiral-odd functions, which, in 
particular, may help in determining transversity.

- T-odd functions are of order $g \pi_{\perp}/Q$. This implies a suppression by at least a 
factor of $Q^{-n}$ in the asymmetries involving $n$ such functions, in accord with data of 
unpolarized Drell-Yan and  with the quark-quark-gluon correlations. Our prediction could also 
be compared with data of incoming HERMES, COMPASS and CLAS experiments.

- Nontrivial twist-2 and twist-3 contributions to the combination $g_1+g_2$ come only from 
the quark-gluon anomalous coupling in the singlet axial current or from nonperturbative 
fluctuations of the nucleon.

\end{document}